# High-Temperature Ultra-Broad UV-MIR High-Efficiency Absorber Based on Double Ring-Shaped Titanium Nitride Resonators

**Shun Cao[1], Yi Jin[1], Hongguang Dong[1,2], Tingbiao Guo[1], Zhenchao Liu[1], Jinlong He[3,#] and Sailing He[1,2,4*]**

[1]*Centre for Optical and Electromagnetic Research, National Engineering Research Center for Optical Instruments, Zhejiang University, Hangzhou 310058, China*

[2]*Ningbo Research Institute, Zhejiang University, Ningbo 315100, China*

[3]*College of Modern Science and Technology, China Jiliang University, Hangzhou, 310018, China*

[4]*Department of Electromagnetic Engineering, School of Electrical Engineering, KTH Royal Institute of Technology, S-100 44 Stockholm, Sweden*

[#]jlhe@cjlu.edu.cn

[*]sailing@kth.se

## Abstract

An ultrabroad absorber based on double-ring-shaped titanium nitride (TiN) nanoresonators, which can work in high temperatures, is proposed and numerically studied. The absorber with some optimal parameters exhibits an averaged absorption of 94.6% in the range of 200 - 4000 nm (from ultraviolet to mid-infrared) and a band from 200 - 3518 nm having an absorption > 90%. We have demonstrated in detail the physical mechanisms of the ultra-broad absorption, including the dielectric lossy property of TiN material itself in shorter wavelengths and plasmonic resonances caused by the metallic property of TiN nano-resonators in longer wavelengths. In addition, the absorber shows polarization independent and wide-angle acceptance. Another absorber with double TiN nano-rings of different heights has flatter and higher absorption efficiency (more than 95% absorption) at 200-2860 nm waveband. These properties make the proposed absorbers based on TiN has great potentials in many applications, such as light trapping, photovoltaics, thermal emitters.

## 1. Introduction

In the last decades, the perfect absorbers based on metamaterials have been widely proposed and studied from visible to terahertz domain [1-5], for many potential applications, such as infrared thermal detection [6, 7], energy harvesting [8-10]. Various structures made of noble metals, like gold and silver are widely investigated for the perfect absorbers [1, 11]. However, these perfect absorbers consisting of noble metals suffer from low melting points (about 1000 ℃), which limit their use in thermo-photovoltaic applications. Therefore, it is necessary to study the absorbers based on refractory metals with high melting points [12-15].

Due to its extraordinary properties, such as compatibility with the CMOS processes, high melting point of 2930 ℃, high temperature durability and large loss in the visible range, TiN is a good candidate for the design of prefect absorbers to be used in high-temperature thermo-photovoltaic systems [16, 17]. In addition, TiN possesses the property of the transition from dielectrics to metal at the wavelength of 425 nm. By virtue of this feature, the broadband absorbers based on the TiN materials can be designed. The TiN based perfect absorbers, including square-ring aray, nanodisk array were firstly studied with high absorption limited in the visible regime [16, 17]. More complicated TiN nanostructures, such as TiN 3D-truncated nanopillars and nanocones were proposed to realize broad absorption in the visible and near-infrared (NIR) regions [18, 19]. Then, some absorbers were proposed to efficiently absorb the incident waves in the ultraviolet to NIR ranges [20-22]. Wu *et al* proposed large-area and ultrathin metasurface based on TiN realizing strong ultrabroad absorption from ultraviolet to NIR, however, the average absorption was only 85% [23]. Mehrabi *et al* demonstrated an absorber based on cross-shaped TiN resonators with >90% absorption in the wavelength range of 200 - 2500 nm [24]. The bandwidth can be further extended to mid-infrared (MIR) by adding more TiN resonators into the unit cells.

In this work, we proposed an ultra-broad absorber based on metal-insulator-metal (MIM) configuration that consists of double-ring shaped TiN arrays located on $Al_2O_3$/TiN films. This absorber with total thickness of 490 nm has more than 90% absorption in the wavelength range of 200-3518 nm. The physical mechanisms of the ultra-broad absorption are explained by analyzing the electric field, power flow and power loss distributions. The effects of different geometric parameters on absorption are also investigated. Finally, another absorber was also designed to make the efficiency of the TiN absorber flatter and higher (over 95%) over an ultra-broad band.

## 2. Structure and simulation methods

Figure 1(a) shows the unit cell of the proposed TiN-based absorber (Absorber A), which consists of three layers: a double ring-shpaed TiN resonator, a middle $Al_2O_3$ (melting point of 2000 ℃) spacer layer and a TiN ground plane. The thickness of the ground plane is $t_1$ = 100 nm. The thickness of the $Al_2O_3$ film and topmost TiN resonator are chosen as $t_2$ = 70 nm and $t_3$ = 320 nm, respectively. The double TiN nano-ring arrays are periodically arranged along both the x and y directions with P = 460 nm. The top view of the unit cell is depicted in Fig. 1(b). The width $w_1$ and outer-radius $r_1$ of TiN inner nano-ring are $w_1$ = 60 nm and $r_1$ = 85 nm, respectively. And the inner-radius $r_2$ and outer-radius $r_3$ of TiN outer nano-ring are $r_2$ = 150 nm and $r_3$ = 170 nm, respectively.

The optical responses, including reflection and transmission spectra of the designed TiN absorbers in this work, are simulated using finite-difference time-domain (FDTD) method (Lumerical FDTD Solutions). In the regions surrounding the absorbers, finer meshes with cubes of 1 × 1 × 1 nm³ are applied to ensure the accuracy and stability of FDTD calculations. Coarser meshes are used elsewhere. The refractive index of the TiN and Al$_2$O$_3$ are obtained from the experimental data by Palik [25]. The periodic boundary conditions (PBC) are employed along x and y directions and the perfectly matched layer (PML) boundary condition is employed in the z axis. The absorption A is extracted by A = 1 - R - T, where R and T are the corresponding reflection and transmission, respectively.

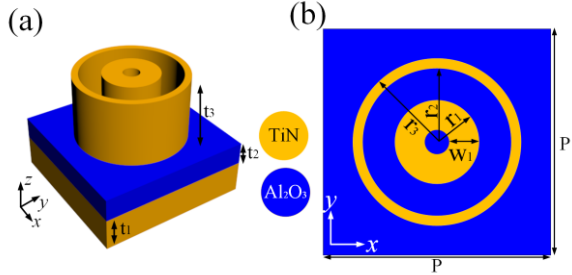

Figure 1. (a) Schematic diagram for the unit cell of the proposed Absorber A. (b) Top view of the structure.

## 3. Results and discussions

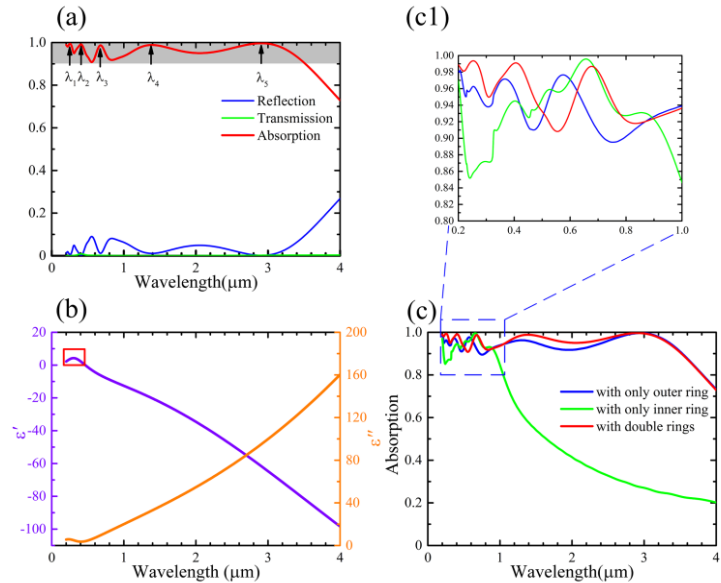

Figure 2. (a) The simulated reflection, transmission and absorption spectra of the proposed Absorber A under x-polarized normal incidence. (b) The permittivity of TiN. ε' (ε'') denotes the real (imaginary) part. (c) A comparison of absorption spectrum of the proposed Absorber A with double rings and the absorbers with only one ring. (c1) An enlarged view of the comparison in (c) from 200 nm to 1000 nm.

Figure 2(a) shows the simulated reflection (blue curve), transmission (green curve) and absorption (red curve) spectra of the Absorber A with a double-ring resonator under the normally incident x-polarized light. One can see from the absorption curve in Fig. 2(a) that this absorber can absorb over 90% of the incident waves from 200 nm to 3518 nm with an average absorption 96.4%, which shows the absorption bandwidth is more than 3300 nm. The average absorption in the range of 200-4000 nm is 94.6%. To the best of our knowledge, this bandwidth is the largest when compared to other TiN MIM absorbers [16, 17, 19, 21, 22, 24]. For the absorption, there are five peaks: $\lambda_1$ (253 nm), $\lambda_2$ (406 nm), $\lambda_3$ (678 nm), $\lambda_4$ (1388 nm) and $\lambda_5$ (2899 nm), which have almost near-unity absorption. The permittivity of TiN is shown in Fig. 2(b). It is worth noting that the real part $\varepsilon'$ is positive (exhibits dielectric property) when $\lambda <$ 465 nm, as showing in red region in Fig. 2(b), while it becomes negative (shows metallic property) when $\lambda >$ 465 nm. It can be seen that two peaks fall into the dielectric range and three peaks into the metallic range. The absorption curves of the absorbers with only one ring resonator under the same parameters are also calculated and given in Fig. 2(c) to show the effects of different rings on the absorption. The detailed absorption from 200 to 1000 nm is shown in Fig. 2(c1). From these figures, it can be found that the outer ring plays an important role in the high absorption for longer wavelengths. The absorption curves for the double-ring and only outer-ring absorbers show almost the same trends and exhibit nearly the same values from 2800 to 4000 nm. In addition, both rings have effects on the high absorption from ultraviolet to visible wavelengths. In order to demonstrate the effects of the patterned top TiN resonators on the absorption, the absorption of two planar TiN MIM absorbers are calculated and shown in Fig. 3. Both planar absorbers consist of TiN/$Al_2O_3$/TiN films. One (Absorber C1) has optimal thickness (the thickness of top planar TiN film is 10 nm and other parameters are same with proposed TiN absorber) to achieve high absorption and the other (Absorber C2) has the same thickness as the proposed TiN absorber. Absorber C2 has the worst absorption among these three absorbers. After thickness optimization, Absorber C1 exhibits 97.7% absorption at 568 nm. It can be seen that the absorption is greatly improved in broadband when the top TiN film is patterned.

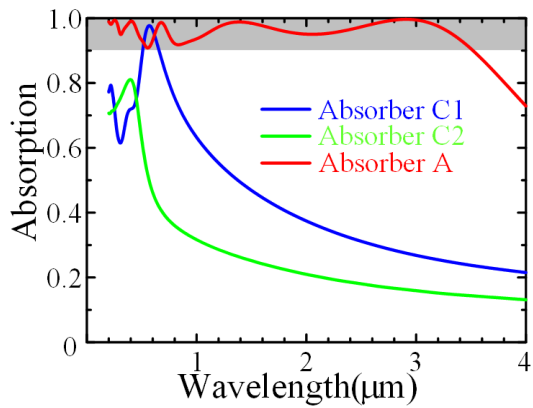

Figure 3. A comparison of absorption spectra of the Absorber A and two planar MIM absorbers.

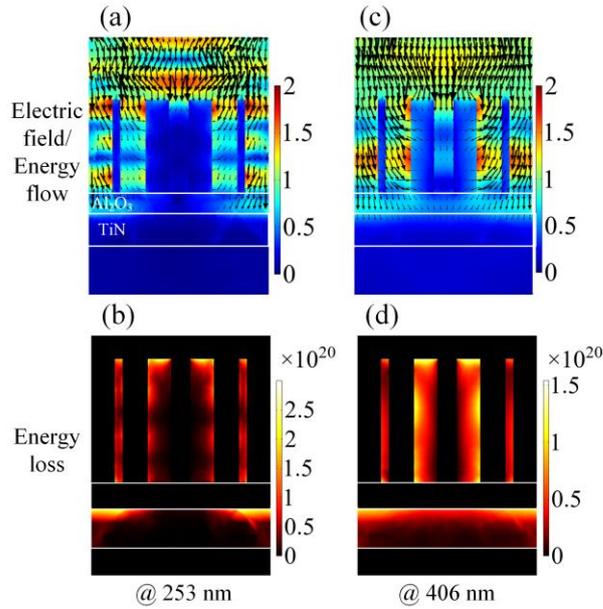

*Figure 4. The electric field (|E/E₀|) (color maps in the first row), energy flow (black arrow maps in the first row) and energy loss (color maps in the second row) distributions of the absorber at 253 nm and 406 nm under x-polarized normal incidence in the x–z plane.*

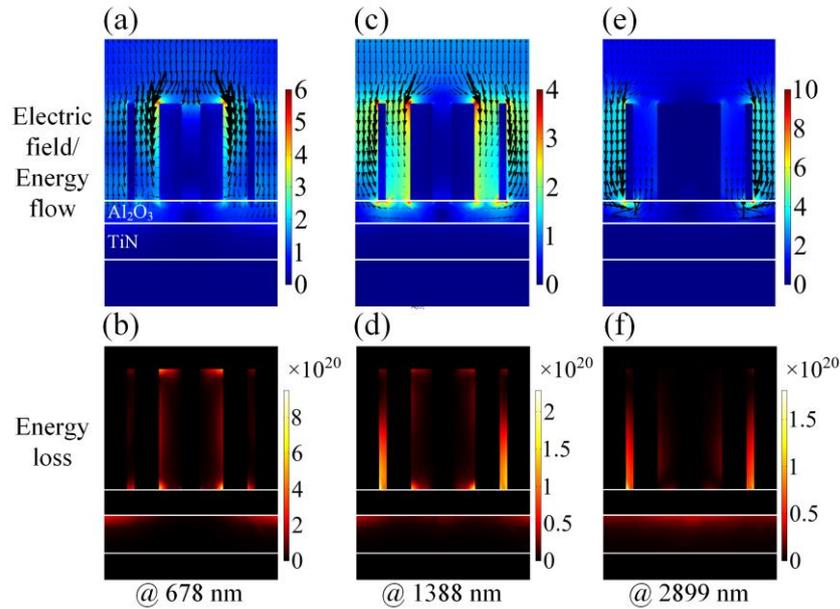

*Figure 5. The electric field (|E/E₀|) (color maps in the first row), energy flow (black arrow maps in the first row) and energy loss (color maps in the second row) distributions of the absorber at 678 nm, 1388 nm and 2899 nm under x-polarized normal incidence in the x–z plane.*

To better reveal the mechanism for the contribution of the ultra-broad absorption of the proposed absorber, distributions for the electric field enhancement ($|E/E_0|$), energy flow and energy loss in the x-z plane at five peaks (253, 406, 678, 1388 and 2899 nm) under normal incidence are simulated and the results are shown in Figs. 4 and 5. For short wavelength peaks (253 and 406 nm), the TiN exhibits high lossy dielectric property. From Figs. 4(a) and 4(b), one sees that the energy can penetrate inside the top TiN nano-rings and bottom TiN film. The top TiN nano-rings act as dielectric electric resonators, which localize the electric field at the sidewalls and top corners of the rings. From the energy flow (the black arrows) and energy loss distributions in Fig. 4, one sees that the energy is absorbed at where it flows due to the large loss of the TiN and enhanced electric field. The length of the arrows are proportional to the light intensity. Specifically, at wavelength peak 253 nm, the electric field is concentrated at the outer sidewall of the outer-ring and the top corners of the inner-ring. Thus, the energy loss at these locations are the largest. However, at wavelength peak 406 nm, the electric field is localized at the inner sidewall of the outer-ring and outer-sidewall of the inner-ring, where energy loss are the greatest. As shown in Figs. 4(a) and 4(b), the incident light can penetrate deep into the bottom TiN film. Therefore, it can be concluded that the formation of highly lossy dielectric and electric resonators is a key to realize high absorption. For long wavelength peaks (678, 1388 and 2899 nm), as the TiN exhibits metallic property, the electric field is strongly localized at the top corners of the TiN nano-rings and the interface between TiN nano-rings and $Al_2O_3$ film, suggesting the excitation of the LSPR and Fabry-Perot resonances as shown in Figs. 5(a), 5(c) and 5(e). At wavelength peak 678 nm, the localized electric field is concentrated at the top corner of the TiN inner-ring and the interface between the TiN inner-ring and $Al_2O_3$ film. For wavelength peak 1388 nm, the electric field is strongly confined at the positions of the top corners of both TiN nano-rings and the interface between TiN resonators and $Al_2O_3$ film. However, the localized field (hot area) will move to the TiN outer-ring when the wavelength becomes longer (cf. Figs. 5(c) and 5(e)). Therefore, consistently matches the conclusions from Fig. 2(c), the outer-ring plays an important role for the high absorption at longer wavelengths, while both rings have effects on the high absorption from ultraviolet to visible wavelengths. At wavelength peak 2899 nm, the incident power flows into the gaps between the adjacent outer-rings and gap resonances are formed. From Figs. 5(b), 5(d) and 5(f), the bottom TiN film also absorb some portion of incident waves. As the thickness of the TiN film is 100 nm (greater than the skin depth of TiN), the transmission is nearly 0 at the longer wavelengths (can also be seen from Fig. 2(a)). In short, the ultra-broad absorption of the proposed absorber can be attributed to the dielectric lossy property of TiN itself and excited electric resonance in shorter wavelengths and the plasmonic resonances caused by the TiN nano-resonators in longer wavelengths.

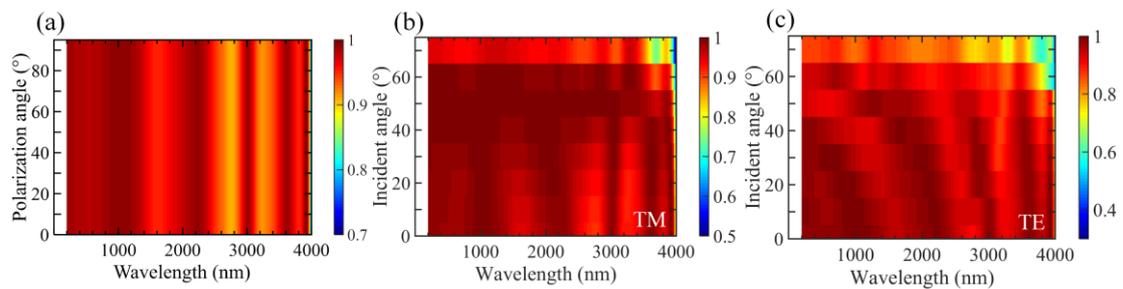

Figure 6. The simulated absorption spectra under different (a) polarization angles (0-90°) anddifferent incident angles (0-70°) for (b) TM-polarized and (c) TE-polarized light.

The effects of different polarization directions and the oblique incidences on the absorption of the absorber are further studied. With the polarization angle from 0° to 90°, the normal incident wave changes from x-polarized to y-polarized. As shown in Fig. 6(a), the simulated absorption remains unchanged, due to the geometric symmetry of TiN double-ring shaped resonators. Therefore, our proposed ultra-broad absorber is polarization independent for normal incidence. However, for oblique incidence, the absorption change will be different for different polarizations. Figures 6(b) and 6(c) shows the absorption spectrum under the TM-polarized and TE-polarized incident with incident angles from 0° to 70°, respectively. As can be seen from Fig. 6(c), for TM-polarized incidence, the absorber maintains high absorption when the incident angle increases to 60°. When the incident angle is 70°, the average absorption reduces to 77.9%. However, the absorber still has an 92.4% average absorption in the ultraviolet and visible ranges (200-800 nm). For TE-polarized, from Fig. 6(c), the absorption degrades with the incident angle increases. The absorber maintains high absorption when the incident angle reaches 50°, with 89.3% average absorption at the wavelength of 200-3518 nm. As the incident angle increases to 60° and 70°, the average absorption reduces to 75.3% and 68.8%, respectively. However, the absorber under the incident angle of 70° still has an 81.4% average absorption in the ultraviolet and visible ranges.

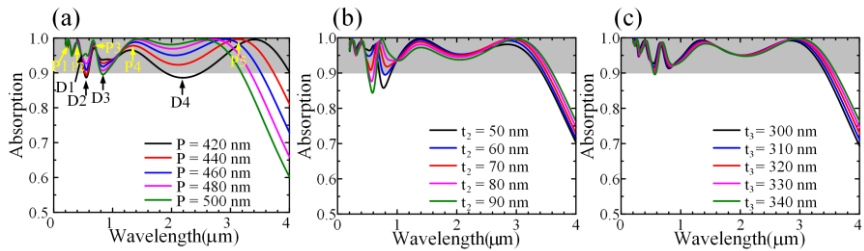

Figure 7. The absorption spectra of the Absorber A for different values of (a) period P, (b) thickness of $Al_2O_3$ film $t_2$ and (c) thickness of top TiN resonantors $t_3$. When one parameter changes, the other parameters of the absorber are kept at the optimum values.

The effects of different geometric values of this ultra-broad Absorber A on the absorption are also investigated. As can be seen from Fig. 7(a), the absorption curve shows different changes for different wavelength regions. There are five peaks (P1, P2, P3, P4 and P5) and four dips (D1, D2, D3 and D4) in all absorption spectrum. With the period increases, the P1, P2 and D1 have little change, which means the absorption in the ultraviolet is almost unchanged. However, the absorption values at the D2, D4, P3 and P4 gradually increase with the period increase and only the wavelengths of P4 and D4 show obviously redshift and blueshift, respectively. For D3, the absorption will have a decrease as the period increases, while the resonance wavelength almost keeps the same. Although absorption value almost has no change, the wavelength of P5 shows apparently blue-shift when period of the absorber increases. This means the period has obviously influences on the resonance P5, which can be also explained by Fig. 5(e). Figure 7(b) shows the effects of different thickness of $Al_2O_3$ film on the absorption. It can be found that all the peaks and dips show different degrees of red-shifts with the $t_2$ increases. In addition, the absorption values of D2 and D3 are most dramatic changed. However, all the peaks and dips of the absorption curves in Fig. 7(c) show red-shifts with almost unchanged absorption values when the thickness of top TiN resonators increases. After the above parameter analysis, the bandwidth of the absorption >90% reaches the maximum when P = 460 nm, $t_2$ = 70 nm and $t_3$ = 320 nm.

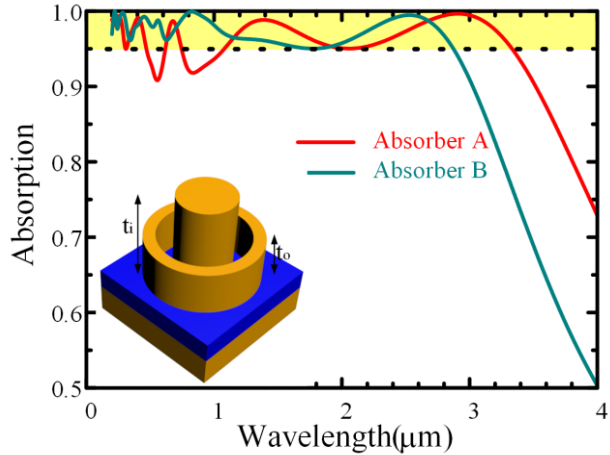

Figure 8. A comparison of absorption spectra of Absorber A and Absorber B which has different heights for the double rings. The inset is the schematic diagram of Absorber B, where the double rings of the top TiN resonator have different heights $t_o$ and $t_i$. The other structural parameters are the same as those in Figure 1. The absorption is over 95% in a broad band from UV to MIR.

From Figs. 2(a) and 2(c1), it can be found that the absorption in some waveband is lower than 95% in the visible and NIR range. In order to broaden the bandwidth (the absorption >95%), another absorber with double rings of different heights, as shown in the inset of Fig. 8, is also proposed. The parameters of this absorber are $t_1$ = 100 nm, $t_2$ = 90 nm, $t_o$ = 170 nm, $t_i$ = 380 nm, $r_1$ = 95 nm, $r_2$ = 160 nm, $r_3$ = 200 nm, and P = 520 nm. From Fig. 8, one can see that Absorber B has flatter and higher absorption efficiency. It has more than 95% absorption in 200-2860 nm waveband (while Absorber A has more than 95% absorption only in 200-470 nm, 618-744 nm and 1081-3033 nm wavebands, and the absorption is below 95% in the important visible range 470-618 nm and NIR range 744-1081nm). Therefore, both absorbers proposed in this paper show great abilities to absorb broadband incident light.

## 4. Conclusion

In summary, we have proposed and numerically demonstrate an ultra-broad MIM absorber, which consists of TiN bottom film, $Al_2O_3$ dielectric film and double-ring shaped TiN resonators, which could be used in high temperatures. This absorber shows >90% absorption at from 200 to 3518 nm with an average absorption 96.4%. The ultra-broad absorption of the proposed absorber is attributed to the dielectric lossy property of TiN itself in shorter wavelengths and the plasmonic resonances caused by the TiN nano-resonators in longer wavelengths. The proposed absorber also shows polarization insensitivity and exhibits high absorption (more than 90%) for angles from 0° to 50°. By utilizing the double rings with different heights, the abosrber can have more flat and high absorption at 200-2860 nm waveband (more than 95% absorption). The absorbers in this work could be applied into many potential fields, like thermophotovoltaics, light trapping and other optoelectronic devices.

## Acknowledgements

The authors thank Dr. Liu Yang for helpful discussion. This work was partially supported by the National Natural Science Foundation of China (No. 91833303, 11621101, 61774131) and the National Key Research and Development Program of China (No. 2017YFA0205700)